\documentclass[sigconf, nonacm]{acmart}

\usepackage{pvldb}

\usepackage{amsmath}
\usepackage{booktabs}
\usepackage{makecell}
\usepackage{tabularx}
\usepackage{array}
\usepackage{multirow}
\usepackage{float}
\usepackage{xspace}
\usepackage{enumitem}
\usepackage{graphicx}
\usepackage{tikz}
\usetikzlibrary{arrows.meta,positioning,fit}
\usepackage{microtype}
\usepackage{stfloats}

\newcommand{\PaperTitle}{SQL-RewriteBench: A Correctness-Gated and Full-Denominator Benchmark for Statement-Level SQL Rewriting [Experiment, Analysis \& Benchmark]}

\newcommand{\PaperAuthors}{Jiang Long, Tianci Gao, Shiyuan Hao, Haochen Zhang, Shuncheng Liu, and Jiang Zhang}
\newcommand{\ArtifactURL}{https://github.com/SQL-RewriteBench/benchmark}

\newcommand{\NumCases}{180\xspace}
\newcommand{\PGAwareCases}{30\xspace}
\newcommand{\EquivCases}{48\xspace}
\newcommand{\PerfCases}{44\xspace}
\newcommand{\RobustCases}{58\xspace}

\newcommand{\BenchName}{SQL-RewriteBench\xspace}
\newcommand{\CGOQ}{CGOQ\xspace}
\newcommand{\SCS}{SCS\xspace}

\newcommand{\parhead}[1]{\par\noindent\textbf{#1}}
\newcolumntype{Y}{>{\raggedright\arraybackslash}X}
\newcolumntype{L}[1]{>{\raggedright\arraybackslash}p{#1}}
\newcolumntype{C}[1]{>{\centering\arraybackslash}p{#1}}

\renewcommand\vldbdoi{XX.XX/XXX.XX}
\renewcommand\vldbpages{XXX--XXX}
\renewcommand\vldbauthors{\PaperAuthors}
\renewcommand\vldbtitle{\PaperTitle}
\renewcommand\vldbavailabilityurl{\ArtifactURL}

\setlist[itemize]{leftmargin=*,nosep}
\setlist[enumerate]{leftmargin=*,nosep}
\begin{document}
\flushbottom

\title{\PaperTitle}

\author{Jiang Long}
\affiliation{%
  \institution{Zhejiang University}
  \city{Hangzhou}
  \country{China}
}
\additionalaffiliation{%
  \institution{Huawei Company}
  \country{China}
}
\email{longjiang1@zju.edu.cn}

\author{Tianci Gao}
\affiliation{%
  \institution{Huawei Company}
  \city{Hangzhou}
  \country{China}
}
\email{gao.tianci@huawei-partners.com}

\author{Shiyuan Hao}
\affiliation{%
  \institution{Huawei Company}
  \city{Hangzhou}
  \country{China}
}
\email{haoshiyuan@huawei.com}

\author{Haochen Zhang}
\affiliation{%
  \institution{Huawei Company}
  \city{Hangzhou}
  \country{China}
}
\email{zhanghaochen6@huawei-partners.com}

\author{Shuncheng Liu}
\affiliation{%
  \institution{Huawei Company}
  \city{Hangzhou}
  \country{China}
}
\email{liushuncheng1@huawei.com}

\author{Jiang Zhang}
\affiliation{%
  \institution{Huawei Company}
  \city{Hangzhou}
  \country{China}
}
\email{zhangjiang13@huawei.com}

\begin{abstract}
Statement-level SQL rewriting can improve query performance and maintainability without changing the DBMS kernel, but existing benchmarks do not evaluate rewrite methods as deployable systems. We present \textsc{SQL-RewriteBench}, a benchmark that applies correctness gating and full-denominator accounting. Its denominator-explicit metric suite measures Source Acceptance, Generation Rate, Execution Coverage, Result Consistency, UnsafeRewrite Rate, and speedup distribution. It also defines SCS, a deterministic index of static SQL structure, and CGOQ, which scores runtime improvement and structural simplification only after the case-specific Checker Contract is satisfied.

\textsc{SQL-RewriteBench} provides 180 executable Benchmark Instances organized into EQUIV, PERF, ROBUST, and PostgreSQL-aware pools, with SQL, schema metadata, provenance, evidence, and rewrite-opportunity documentation. Across seven representative academic and LLM-based methods, every full-benchmark CGOQ is negative. Existing methods often fail before rewriting, fail result checks, or return correct rewrites that are slower or no better than the input. These results highlight the need for broader input handling, result validation, and benefit-aware rewrite decisions.
\end{abstract}

\maketitle

\vldbtopmatter

\section{Introduction}
\label{sec:introduction}

SQL rewrite is a long-standing database optimization technique: optimizers and external tools transform a SQL statement to expose better plans, simplify expressions, decorrelate subqueries, push predicates, or remove redundant structure~\cite{selinger1979access,pirahesh1992starburst,graefe1993volcano,graefe1995cascades}. Recent systems revisit statement-level rewrite outside the DBMS kernel, using rule search, learned rule selection, retrieval, or language models to generate a replacement SQL statement~\cite{zhou2022learnedrewrite,li2024llmr2,sun2025rbot,song2025quite}. This deployment model has a visible SQL-in/SQL-out contract: a method must accept the input, decide whether to rewrite, emit one Rewritten Query, and submit it to the evaluation DBMS.

Existing benchmarks and tools do not fully capture this rewrite-specific SQL-in/SQL-out contract. DBMS workloads such as TPC-H, TPC-DS, DSB, and JOB measure execution or optimizer behavior~\cite{tpch,tpcds,ding2021dsb,leis2015job}; dialect datasets study translation~\cite{zhou2025parrot}; equivalence systems decide supplied query pairs over supported fragments~\cite{chu2018usemiring,zhou2022spes,ding2023sqlsolver,wang2024qed,he2024verieql}. To our knowledge, no existing resource keeps Source Acceptance Failure, No-Rewrite Decision, invalid generation, DBMS execution failure, result inconsistency, Unsafe Rewrite, speedup distribution, and structural simplification in one explicit denominator. As a result, an average speedup over successful cases can hide many operational failures.

\BenchName addresses this gap. Its first contribution is methodological. We define a denominator-explicit metric suite for deployable SQL rewrite and introduce two script-computable scores. Static SQL Complexity Score (SCS) measures deterministic AST structure; Correctness-Gated Optimization Quality (CGOQ) assigns optimization value only after the case-specific Checker Contract is satisfied. CGOQ prioritizes runtime improvement and gives bounded structural-simplification credit only when runtime is neutral or the slowdown is small; the credit is attenuated as slowdown grows and vanishes at the default 10\% slowdown threshold.

The second contribution is an executable benchmark artifact. The release contains \NumCases Benchmark Instances with documented rewrite opportunities in four pools: EQUIV, PERF, ROBUST, and PostgreSQL-aware (PG-AWARE). Each instance is an Executable Case Package with a Benchmark Input Query, a Reference Rewrite, Base Query provenance, schema metadata, result and plan evidence, a Checker Contract, pool-assignment evidence, and a human-readable rewrite-opportunity account. The Evaluation Harness runs methods, checks results, computes SCS and CGOQ, and emits a full-denominator ledger.

The third contribution is experimental evidence. We evaluate representative academic and LLM-based methods, including two official LLM-R2 demonstration-selection modes. The results do not merely show that particular baselines underperform; they show where representative evaluated rewrite systems fail and what future systems must improve: broader SQL front ends, cleaner rule-executor interfaces, result-aware validation, and more selective optimization policies.

\section{Related Work}
\label{sec:related}

\parhead{Database workloads and SQL corpora.}
TPC-H, TPC-DS, DSB, JOB, and OLTP-Bench provide workloads and DBMS protocols for measuring execution behavior~\cite{tpch,tpcds,ding2021dsb,leis2015job,difallah2013oltpbench}. SQLStorm stresses complex SQL behavior~\cite{sqlstorm}; Spider, BIRD, and complex-SQL datasets characterize structural difficulty for text-to-SQL~\cite{yu2018spider,li2023bird,ma2024complexsql}; PARROT targets dialect translation~\cite{zhou2025parrot}. Cross-dialect translation is outside our task: PG-AWARE fixes PostgreSQL as both the source dialect and the evaluation DBMS and evaluates same-dialect rewriting of PostgreSQL-sensitive constructs. These resources supply Base Queries or structural insights, but they do not provide opportunity-enriched inputs, validated Reference Rewrites, result evidence, pool-assignment rationale, and correctness-gated optimization scores for statement-level rewrite.

\parhead{SQL equivalence and correctness evidence.}
HoTTSQL, Cosette, U-semiring, SMT-based work, SPES, SQLSolver, QED, and VeriEQL develop formal semantics and equivalence-checking techniques~\cite{chu2017hottsql,chu2017cosette,chu2018usemiring,zhou2019smt,zhou2022spes,ding2023sqlsolver,wang2024qed,he2024verieql}. They are valuable when a query pair lies in a supported fragment, but they neither generate rewrites nor define an end-to-end deployment outcome. \BenchName instead uses an execution-based Checker Contract as the operational Correctness Gate and reports that limitation explicitly.

\parhead{SQL rewrite systems.}
LearnedRewrite searches rewrite sequences with Monte Carlo tree search~\cite{zhou2022learnedrewrite}; WeTune and GRewriter discover and verify rewrite rules~\cite{wang2022wetune,jiang2025grewriter}; QueryBooster and SlabCity study middleware and synthesis~\cite{bai2023querybooster,dong2023slabcity}. LLM-R2, R-Bot, QUITE, and GenRewrite introduce language models for rule selection, evidence retrieval, feedback, or direct generation~\cite{li2024llmr2,sun2025rbot,song2025quite,liu2024genrewrite}. These papers evaluate their own systems; \BenchName standardizes the input population, terminal outcomes, denominator, and quality model across methods.

\section{Task Scope and Design Principles}
\label{sec:task}

\subsection{Optimization Value and Task Definition}

The immediate value of SQL optimization is lower latency and fewer timeouts; at workload scale it can also improve throughput and reduce the probability of resource exhaustion. The engineering value of a rewrite can persist even when runtime is unchanged. Fewer nested scopes, less duplicated logic, and clearer relational boundaries reduce the effort required to review, test, modify, and diagnose long-lived analytical SQL. We focus on execution performance and static structural complexity because both can be measured consistently from a submitted statement and its repeated executions. Engine-specific CPU, I/O, memory, and cloud-cost counters are useful operational profiles, but they are not part of the core score because their semantics differ across DBMSs.

A Benchmark Instance consists of a \emph{Benchmark Input Query} and its declared execution context. The context includes the SQL dialect, schema and relevant constraints, validation database, exact-result evidence, and case-specific preservation conditions. A method returns one of two outputs:
\begin{itemize}
  \item a \textbf{Rewritten Query}, a complete SQL statement intended to replace the Benchmark Input Query under the same context; or
  \item a \textbf{No-Rewrite Decision}, indicating that the method elects not to change the input.
\end{itemize}
Returning the input verbatim after normalization, or changing only formatting, comments, or superficial aliases, is a \emph{no-op}. No-Rewrite and no-op are safe outcomes, but neither contributes optimization value.

The main protocol is top-1. A method may submit at most one final Rewritten Query per instance. It must choose the SQL that would be sent to the DBMS; it is not credited for alternative candidates that it did not select as its final output. This rule prevents oracle selection after correctness and runtime outcomes are known.

\subsection{Evaluation Scope}

\BenchName evaluates complete statement-level SQL under a declared source dialect and evaluation DBMS. It does not evaluate an internal physical plan, an index recommendation, or a materialized-view design. PG-AWARE instances test same-dialect rewriting of PostgreSQL-sensitive syntax and behavior; cross-dialect translation is outside the task.

Also outside scope are join-order optimization and cardinality estimation as independent tasks, workload scheduling and caching, physical-design selection, and traces that do not emit a complete SQL statement. These tasks can affect runtime, but they have different inputs and success conditions. Keeping them separate makes the rewrite contract and its failure denominator explicit.

\subsection{Design Goals}

Table~\ref{tab:goals} maps deployment questions to the evidence collected by the benchmark. The layers form a funnel, but headline rates retain the full planned denominator. Consequently, a parser rejection, a No-Rewrite Decision, an execution error, and a result mismatch stay visible as different outcomes.

\begin{table}[t]
\centering
\caption{Deployment-oriented design goals.}
\label{tab:goals}
\small
\begin{tabularx}{\columnwidth}{L{.18\columnwidth}Y Y}
\toprule
Goal & Question & Evidence \\
\midrule
Acceptance & Can the system enter its rewrite pipeline? & Source Acceptance Rate \\
Generation & Does it emit a substantive SQL statement? & Generation and No-Rewrite rates \\
Execution & Can the output run on the evaluation DBMS? & Execution Coverage Rate \\
Correctness & Does it preserve the observed case-specific Checker Contract? & Result Consistency and UnsafeRewrite Rate \\
Optimization & Does it improve runtime, or simplify SQL when runtime is neutral or only slightly slower? & CGOQ, speedup distribution, SCS \\
Inspectability & What opportunity did the output address? & Rewrite Target Coverage and evidence ledger \\
\bottomrule
\end{tabularx}
\end{table}

The distinction between a safe No-Rewrite Decision and a successful optimization is important. A conservative system may decide not to rewrite when the evidence is insufficient. However, because every released Benchmark Instance contains a validated opportunity, a No-Rewrite Decision is a missed opportunity rather than a success. At the other extreme, high generation coverage cannot compensate for an executable statement that fails the case-specific Checker Contract.

\subsection{Terminal Outcome Semantics}

Each method--instance pair receives one terminal status before aggregation. A Source Acceptance Failure occurs when mandatory source-side parsing, validation, catalog resolution, or relational conversion terminates before the method returns an explicit SQL output or No-Rewrite Decision. An internal front-end error followed by a documented fallback and an explicit method decision is counted as accepted. A Generation Failure occurs after Source Acceptance when the method returns neither a complete Rewritten Query nor a recognized No-Rewrite Decision. An Execution Failure is a generated statement rejected by the evaluation DBMS. An Unsafe Rewrite executes but fails the case-specific Checker Contract. A result-consistent output is then classified as Performance Gain, Simplification Gain, Correct-but-Neutral, or Performance Regression by CGOQ. These mutually exclusive states keep operationally different failures from being collapsed into one ``unsuccessful'' bucket and make the denominator auditable.

\section{Benchmark Construction}
\label{sec:construction}
\subsection{Base Queries and Opportunity Enrichment}
\label{sec:base-query-enrichment}

We first build a broad candidate inventory from DSB, TPC-DS, JOB, Calcite and DBMS regression tests, SQLStorm, PARROT, and related SQL corpora~\cite{ding2021dsb,tpcds,leis2015job,begoli2018calcite,sqlstorm,zhou2025parrot}. Each candidate retains its source path, dataset identity, source family, and usage information. We reject incomplete statements, mutation-only SQL, unresolved templates that affect semantics, missing execution context, near duplicates, and queries without an auditable statement-level rewrite objective. The executable release is drawn from a documented inventory containing 39 SQLStorm entries, 60 Calcite test-dataset entries, 34 TPC-DS entries, and 47 DSB entries. The 180 frozen Benchmark Instances evaluated in this paper are validated on PostgreSQL. The broader inventory is retained as provenance and as input to future releases, but only the frozen Executable Case Packages are used for the experimental results reported here.

A query collected from an existing workload does not necessarily expose a measurable statement-level rewrite opportunity. In many production workloads, a SQL statement may be valid and realistic but already close to the form preferred by the evaluation DBMS, or its optimization opportunity may be too small to distinguish from measurement noise. \BenchName therefore uses DBA-guided, pool-specific opportunity enrichment to construct Benchmark Input Queries with explicit and auditable rewrite objectives. The goal is not to create artificial slowdowns. Instead, enrichment makes realistic industrial rewrite opportunities observable, such as delayed predicate evaluation, redundant query-block or CTE boundaries, repeated computation, decorrelatable subqueries, avoidable aggregation-after-join patterns, long-tail nesting, and dialect-sensitive expressions.

The enrichment policy is specialized by pool. EQUIV stresses correctness-sensitive transformations involving NULL behavior, bag semantics, grouping, ordering, set semantics, and outer joins. PERF introduces recoverable performance anti-patterns such as late filtering, repeated aggregation, materialization barriers, and avoidable join inputs. ROBUST adds nested query blocks, CTE chains, windows, set operations, correlated subqueries, and interacting semantic guards. PG-AWARE makes PostgreSQL-sensitive constructs explicit while preserving PostgreSQL behavior. Multiple DBAs designed these strategies and applied them to selected Base Queries. As an external realism check, five DBAs conducted a blind review of a sample without knowing whether each statement was a Base Query or an enriched Benchmark Input Query. They judged the structures and anti-patterns plausible in production ETL, BI, and reporting workloads, particularly for SQL written by less experienced developers. We therefore treat the enrichment intensity as realistic for capability testing, while making no claim about the production frequency of these patterns.

Table~\ref{tab:enrichment-atoms} reports the most frequent transformation atoms. Counts are non-exclusive because one Benchmark Instance may contain several rewrite opportunities. The distribution spans query-boundary structure, repeated work, predicate placement, relational transformations, sort or duplicate-work patterns, and dialect-sensitive canonicalization. No individual atom occurs in more than 51.1\% of the corpus, reducing dependence on a single synthetic pattern. At the broader non-exclusive family level, 142 cases contain repeated-work or reuse opportunities, 93 contain query-boundary structure, 85 contain dialect or canonical-form opportunities, 70 contain predicate or access-path opportunities, 42 contain relational transformations, and 17 contain sort or duplicate-work opportunities. The distribution is intentionally not uniform: PERF emphasizes performance-oriented anti-patterns, ROBUST emphasizes structural breadth, EQUIV emphasizes semantic boundaries, and PG-AWARE emphasizes PostgreSQL-sensitive behavior. The artifact exposes both atom-level and family-level counts for audit.

\begin{table}[t]
\centering
\caption{Frequent opportunity-enrichment atoms in the 180-case release. Counts are non-exclusive.}
\label{tab:enrichment-atoms}
\scriptsize
\setlength{\tabcolsep}{3pt}
\begin{tabular}{lrr@{\hspace{8pt}}lrr}
\toprule
\textbf{Atom} & \textbf{$n$} & \textbf{\%} & \textbf{Atom} & \textbf{$n$} & \textbf{\%} \\
\midrule
Redundant derived table & 92 & 51.1 & Repeated computation & 76 & 42.2 \\
Canonicalization & 67 & 37.2 & Nested query blocks & 66 & 36.7 \\
Shared precomputation & 64 & 35.6 & Delayed predicate & 60 & 33.3 \\
Redundant CTE boundary & 45 & 25.0 & Correlated subquery & 20 & 11.1 \\
Dialect wrapper & 18 & 10.0 & Sort elimination & 16 & 8.9 \\
Set-operation structure & 13 & 7.2 & Window recomputation & 13 & 7.2 \\
\bottomrule
\end{tabular}
\end{table}

Opportunity enrichment is constrained by a case-specific Checker Contract. It may delay predicate evaluation, retain an avoidable query-block or CTE boundary, introduce repeated computation, expose decorrelation, or delay a valid pre-aggregation until after a join. It must not change join predicates, selection semantics, projected attributes, grouping semantics, duplicate multiplicities, NULL behavior, or observable ordering solely to create a runtime difference. Nondeterministic functions and unbounded Cartesian products are excluded. Each enriched Benchmark Input Query is manually reviewed and tested against its Reference Rewrite. The Executable Case Package stores the original statement as \texttt{provenance/base\_query.sql}, the method input as \texttt{sql/benchmark\_input.sql}, and the validated improved statement as \texttt{sql/reference\_rewrite\_01.sql}. The human-readable \texttt{rewrite\_opportunity.md} explains the applied transformation and the semantic boundaries that a valid Rewritten Query must preserve.

Each Benchmark Input Query--Reference Rewrite pair is checked for exact Result Consistency and executed after warm-up for five measured runs, using the median runtime. PERF construction requires at least a 1.10$\times$ speedup and targets 1.30$\times$ when feasible. Other pools may establish optimization value through substantial SCS reduction within the runtime-neutral band, provided that the Reference Rewrite passes the Checker Contract. This construction procedure turns realistic workload SQL into executable, inspectable, and auditable Benchmark Instances while preserving the distinction between the Base Query, the Benchmark Input Query, and the Reference Rewrite.

\subsection{Pool Design and Suitability Scoring}

Table~\ref{tab:pools} summarizes the frozen pool definitions and release sizes. The frozen release assigns \EquivCases cases to EQUIV, \PerfCases to PERF, \RobustCases to ROBUST, and \PGAwareCases to PG-AWARE. Pool membership is determined before method evaluation from the Benchmark Input Query, schema and dialect metadata, rewrite-opportunity evidence, and the documented Pool Suitability Score. PG-AWARE contains cases for which PostgreSQL-sensitive syntax or behavior is the dominant evaluation question.

\begin{table*}[t]
\centering
\caption{Pool objectives, release sizes, and admission evidence.}
\label{tab:pools}
\small
\begin{tabularx}{\textwidth}{L{.09\textwidth}C{.07\textwidth}L{.22\textwidth}Y Y}
\toprule
Pool & Cases & Dominant question & Quantifiable signals & Primary value \\
\midrule
EQUIV & 48 & Can a method preserve the result at semantic boundaries? & NULL, duplicates, outer joins, set operations, aggregation, casts, ordering, and declared guards & Result-preservation coverage and unsafe-rewrite detection \\
PERF & 44 & Can a method deliver correctness-gated execution improvement? & Delayed filters, repeated work, decorrelation, join-input reduction, aggregation movement, speedup, and plan evidence & Deployable performance optimization \\
ROBUST & 58 & Can a method handle long-tail structural interactions? & Deep nesting, CTE chains, correlation, windows, set operations, complex Boolean logic, and feature interactions & Front-end and rewrite robustness \\
PG-AWARE & 30 & Can a method preserve PostgreSQL-sensitive behavior within the same dialect? & PostgreSQL-sensitive functions, casts, date/time and type behavior, same-dialect execution, and result evidence & Same-dialect, PostgreSQL-aware rewrite \\
\bottomrule
\end{tabularx}
\end{table*}

Pool assignment is automatic but auditable. Let $\operatorname{sat}(n;k)=\min(1,n/k)$. For pool $p$, four normalized evidence components $c_{p,j}(q)\in[0,1]$ define $\operatorname{PoolScore}_p(q)=25\sum_{j=1}^{4}c_{p,j}(q)$.
EQUIV uses opportunity type, semantic hazards, guards, and boundary structure; PERF uses performance opportunities, measured speedup, plan evidence, and performance-relevant structure; ROBUST uses normalized SCS, long-tail features, interactions, and semantic hazards; PG-AWARE uses PostgreSQL-sensitive features, dialect risks, same-dialect validation, and a statement-level opportunity. Automatic admission requires a top score of at least 60 and a margin of at least 5. Borderline cases are reviewed, and \path{case_info.md} records all four scores, recommendation, review status, final assignment, and rationale. Pool Suitability Scores are construction evidence only and do not enter method ranking.

\subsection{Executable Case Package}
\label{sec:case-package}

A SQL file alone is not a reproducible Benchmark Instance. Table~\ref{tab:package} lists the core release-facing files. Each immutable Executable Case Package contains the Benchmark Input Query, validated Reference Rewrite, Base Query, schema profile, canonical result evidence, and execution-plan evidence. \path{case_info.md} records identity, Pool Suitability Scores, assignment basis, source provenance, evaluation environment, referenced tables, validation status, runtime measurements, SCS values, and plan status. \path{rewrite_opportunity.md} explains the case value, concrete transformation, static-feature changes, and semantic boundaries that every valid rewrite must preserve.

\begin{table}[t]
\centering
\caption{Core contents of an Executable Case Package.}
\label{tab:package}
\small
\begin{tabularx}{\columnwidth}{L{.39\columnwidth}Y}
\toprule
Path & Role \\
\midrule
\path{case_info.md} & Identity, assignment evidence, validation summary, lineage, schema, and provenance \\
\path{rewrite_opportunity.md} & Rewrite value, transformation, feature deltas, and semantic boundaries \\
\path{sql/benchmark_input.sql} & Opportunity-enriched SQL submitted to methods \\
\path{sql/reference_rewrite_01.sql} & Validated Reference Rewrite and CGOQ certification evidence \\
\path{provenance/base_query.sql} & Source-aligned Base Query \\
\path{schema/schema_profile.yaml} & Tables, columns, types, constraints, and sample rows \\
\path{evidence/*} & Exact-result and plan evidence for the SQL pair \\
\bottomrule
\end{tabularx}
\end{table}

A method receives the Benchmark Input Query and permitted schema context, but not the Reference Rewrite or validation evidence. Generated SQL, logs, plans, and metrics are stored outside the package. This separation fixes the case-specific Checker Contract across methods while preventing reference imitation. CSV and JSON release ledgers expose pool composition and review decisions programmatically.

\section{Evaluation Methodology}
\label{sec:metrics}

\subsection{Metric Design Principles and Correctness Gate}
\label{sec:metric-principles}

A SQL rewrite method can fail before it produces any useful optimization. It may fail to accept the Benchmark Input Query, return a No-Rewrite Decision, emit an invalid Rewritten Query, produce an executable query that fails the case-specific Checker Contract, or generate a result-consistent rewrite with no measurable optimization value. Reporting only successfully optimized queries would hide these outcomes and overstate end-to-end effectiveness. The metric suite in Table~\ref{tab:metrics} is therefore designed around five principles.

\parhead{Full-denominator accounting.}
All headline rate metrics use the complete set of Benchmark Instances as their denominator. Source Acceptance Failures, No-Rewrite Decisions, generation failures, execution failures, and result-inconsistent rewrites stay visible in the reported results. Performance-distribution metrics, such as GM Speedup Ratio, are computed only over result-consistent queries with valid runtime measurements; their support size is reported explicitly.

\parhead{Stage-separated measurement.}
Source Acceptance, rewrite generation, DBMS execution, Result Consistency, and optimization quality answer different questions and are measured separately. A low Generation Rate may result from Source Acceptance Failures or frequent No-Rewrite Decisions, whereas a high UnsafeRewrite Rate indicates that the method emitted executable but result-inconsistent queries. Separating these stages allows the benchmark to distinguish front-end limitations, conservative behavior, execution failures, and unsafe rewrites.

\parhead{Correctness-gated optimization.}
A Rewritten Query is evaluated for runtime improvement or structural simplification only after it executes successfully and satisfies the case-specific Checker Contract. A faster query receives no optimization credit if it changes the validated result. This principle is also used by CGOQ: optimization quality is defined only after the Correctness Gate is passed.

\parhead{Risk-aware outcome accounting.}
A No-Rewrite Decision is a safe outcome but provides no optimization value. A Source Acceptance Failure or an execution failure reduces coverage. An executable Rewritten Query that fails the Checker Contract is recorded as an Unsafe Rewrite. These outcomes are reported separately because they have different implications for deployment.

\parhead{Reference-independent evaluation.}
Methods are not evaluated by textual or structural similarity to the Reference Rewrite. Any Rewritten Query that satisfies the same Checker Contract is eligible for runtime, SCS, and CGOQ evaluation. The Reference Rewrite is a calibration point, not the only acceptable output.

\begin{table*}[t]
\centering
\caption{Main metrics. Arrows indicate the preferred direction.}
\label{tab:metrics}
\small
\setlength{\tabcolsep}{4pt}
\renewcommand{\arraystretch}{1.08}
\begin{tabularx}{\textwidth}{L{.20\textwidth}L{.22\textwidth}Y}
\toprule
\textbf{Metric} & \textbf{Formula} & \textbf{Interpretation} \\
\midrule
Source Acceptance Rate $\uparrow$ & $|\mathcal{A}_m|/N$ & Fraction that reaches an explicit rewrite decision after mandatory source-side processing or a documented fallback path. \\
Generation Rate $\uparrow$ & $|\mathcal{G}_m|/N$ & Fraction for which the method emits a non-no-op Rewritten Query. \\
No-Rewrite Decision Rate $\downarrow$ & $|\mathcal{R}_m|/N$ & Safe decisions that deliver no optimization value. \\
Execution Coverage Rate $\uparrow$ & $|\mathcal{E}_m|/N$ & Fraction whose Rewritten Query executes on the declared source dialect and evaluation DBMS. \\
Result Consistency Rate $\uparrow$ & $|\mathcal{X}_m|/N$ & Fraction whose Rewritten Query satisfies the case-specific Checker Contract. \\
UnsafeRewrite Rate $\downarrow$ & $|\mathcal{U}_m|/N$ & Executable outputs that fail the Checker Contract. \\
GM Speedup Ratio $\uparrow$ & $\exp(|\mathcal{T}_m|^{-1}\sum_{i\in\mathcal{T}_m}\ln s_i)$ & Geometric mean speedup over result-consistent, timed rewrites. \\
CGOQ@N $\uparrow$ & $N^{-1}\sum_{i\in V_m} c_{m,i}$ & Full-denominator correctness-gated optimization quality reported with coverage and safety rates. \\
\bottomrule
\end{tabularx}
\end{table*}

Let $\mathcal{I}$ denote the complete set of Benchmark Instances and let $N=|\mathcal{I}|$. For method $m$, we use $\mathcal{A}_m$, $\mathcal{R}_m$, $\mathcal{G}_m$, $\mathcal{E}_m$, $\mathcal{X}_m$, $\mathcal{U}_m$, and $\mathcal{T}_m$ to denote, respectively, accepted instances, No-Rewrite Decisions, non-no-op generations, executable rewrites, result-consistent rewrites, unsafe rewrites, and timed result-consistent rewrites. Here $\mathcal{U}_m=\mathcal{E}_m-\mathcal{X}_m$.

The main rate metrics divide the corresponding set sizes by $N$, as shown in Table~\ref{tab:metrics}. Conditional rates, such as $|\mathcal{E}_m|/|\mathcal{G}_m|$, may be reported for failure analysis, but they are not used for leaderboard ranking. Such rates can favor conservative methods that attempt only a small and easy subset of the benchmark.

\parhead{Case-specific Checker Contract.}
The Checker Contract defines the observable properties that a valid Rewritten Query must preserve. By default, query results are compared under bag (multiset) semantics: duplicate multiplicities are preserved, while row order is ignored unless it is observable from the Benchmark Input Query, such as through an \texttt{ORDER BY} combined with \texttt{LIMIT}. NULL values are compared as result cells. Exact numeric types require exact equality, whereas any tolerance for floating-point values must be declared by the Benchmark Instance. Output arity and compatible declared types must match, and the Rewritten Query must execute under the declared source dialect on the evaluation DBMS. A Benchmark Instance may strengthen these requirements but may not remove an observable property of the Benchmark Input Query.

\parhead{Result Consistency rather than formal Semantic Equivalence.}
\BenchName does not claim to establish formal Semantic Equivalence for every Rewritten Query. Existing SQL equivalence verifiers support restricted SQL fragments and require parser, normalization, or relational-conversion pipelines that frequently reject complex, deeply nested, or dialect-specific SQL. In our integration study, the available verifier toolchains could not accept a substantial fraction of the Benchmark Input Queries. Applying formal verification only to the supported subset would make the Correctness Gate query-dependent and would bias the evaluation toward simpler SQL.

We therefore use the case-specific Checker Contract to establish Result Consistency over the packaged schema and data. This execution-based check does not prove that two queries are equivalent over every legal database instance. We also ran SQLSolver and VeriEQL over all reference pairs as supplementary evidence. SQLSolver proved 14 pairs equivalent, returned 115 UNKNOWN, timed out on 51; VeriEQL proved 51 pairs equivalent but returned ERROR on 129 unsupported pairs.  The distinction is explicit throughout \BenchName: Result Consistency is the operational Correctness Gate used by the benchmark, whereas formal Semantic Equivalence is outside the current evaluation scope. The Executable Case Package records the result-comparison policy, schema assumptions, and semantic boundaries needed to interpret this limitation.

The rate metrics and CGOQ provide complementary views of a rewrite method. The rate metrics identify where the method loses coverage or correctness, whereas CGOQ measures the optimization quality of Rewritten Queries that pass the Correctness Gate. CGOQ must therefore be interpreted together with Source Acceptance Rate, No-Rewrite Decision Rate, Generation Rate, Execution Coverage Rate, Result Consistency Rate, and UnsafeRewrite Rate.

\subsection{Static SQL Complexity Score}
\label{sec:scs}

\SCS is a deterministic structural score for this benchmark. It is not an optimizer cost, a semantic-difficulty theorem, or a direct measure of human comprehension. Its role is to make structural changes comparable and script-computable under a frozen parser, source dialect, and feature normalization.

The implementation parses each statement with a dialect-aware AST parser and extracts six feature groups (Table~\ref{tab:scs}). Let $x_f(q)$ be the raw count for feature $f$. We fit a cap only on the frozen Benchmark Input corpus:
\begin{equation}
\begin{aligned}
\tau_f&=\max\left(1,Q_{0.95}\{x_f(q):q\in I\}\right),\\
z_f(q)&=\min\left(1,\frac{\ln(1+x_f(q))}{\ln(1+\tau_f)}\right).
\end{aligned}
\label{eq:scs-normalization}
\end{equation}
For group $g$ with feature set $F_g$,
\begin{equation}
D_g(q)=\frac{1}{|F_g|}\sum_{f\in F_g}z_f(q),
\qquad
\operatorname{SCS}(q)=\frac{100}{6}\sum_{g=1}^{6}D_g(q).
\label{eq:scs-score}
\end{equation}
All six groups and all features within a group are equally weighted after normalization and saturation. This is a transparent, method-independent default rather than a claim that all dimensions impose equal cognitive or optimization cost; the full count vector supports alternative weighting in sensitivity analysis.

\parhead{Runtime association and weighting sensitivity.}
Across the 180 Benchmark Input Queries, SCS has a moderate Spearman association with median runtime ($\rho=0.389$, $p<10^{-7}$), but the within-pool correlations differ: 0.482 for EQUIV, 0.160 for PERF, 0.044 for ROBUST, and 0.222 for PG-AWARE. These heterogeneous correlations support interpreting SCS as a structural stratification measure rather than a runtime surrogate. We also recomputed SCS with topology/nesting, predicate/expression, and analytics/advanced dimensions doubled and renormalized. The method order by CGOQ@N was unchanged under all four schemes, and the largest absolute score shift was below 0.67 points. Table~\ref{tab:scs-validation} summarizes these checks.

\begin{table}[t]
\centering
\caption{SCS dimensions and automatically extracted features.}
\label{tab:scs}
\small
\begin{tabularx}{\columnwidth}{L{.35\columnwidth}Y}
\toprule
Dimension & Representative features \\
\midrule
Relational topology & Table references, joins, outer joins, set operations \\
Nesting/dependency & SELECT blocks, maximum depth, CTEs, CTE-chain depth, correlation \\
Predicate logic & Predicate atoms, Boolean depth, OR branches, CASE expressions \\
Aggregation/analytics & Aggregate calls, grouping keys, HAVING, windows, DISTINCT \\
Expression/projection & Projected expressions, scalar functions, casts, arithmetic \\
Advanced constructs & ORDER/LIMIT, lateral/unnest, recursion, JSON/array/map operations \\
\bottomrule
\end{tabularx}
\end{table}

\begin{table}[t]
\centering
\caption{SCS validation and weighting sensitivity.}
\label{tab:scs-validation}
\scriptsize
\setlength{\tabcolsep}{3.0pt}
\begin{tabularx}{\columnwidth}{L{.38\columnwidth}Y}
\toprule
Check & Result \\
\midrule
Overall SCS--runtime association & Spearman $\rho=0.389$, $p<10^{-7}$ \\
Within-pool $\rho$ & EQUIV 0.482; PERF 0.160; ROBUST 0.044; PG-AWARE 0.222 \\
Alternative dimension weights & Method order unchanged across four schemes \\
Maximum CGOQ@N shift & $<0.67$ points \\
\bottomrule
\end{tabularx}
\end{table}

\parhead{Construct validity.}
SCS is a deterministic index of static SQL structure, not a model of DBMS execution cost or a validated measure of human maintainability. Equal dimension weights provide a transparent, method-independent default after per-dimension normalization and saturation; alternative weight configurations do not change the evaluated method order. As convergent evidence, among the 154 Reference Rewrite pairs with more than 10\% SCS reduction, the reduction is positively associated with lower root plan cost ($\rho=0.532$) and fewer physical plan operators ($\rho=0.693$). These associations are strongest in PERF and weaker in EQUIV and ROBUST. We therefore interpret SCS reduction as a proxy for structural simplification, while leaving the relationship between SCS and human review or maintenance effort to a future user study.

Comments, formatting, and superficial alias changes are normalized before counting. The same frozen $\tau_f$ values are used for Benchmark Inputs, Reference Rewrites, and method outputs. If the declared dialect cannot be parsed, the result is \texttt{PARSE\_ERROR}; the script does not manufacture a pseudo-precise score from regular expressions. CGOQ may still use its performance branch in this case, but not the structural-simplification branch. The artifact emits the aggregate SCS, six group scores, raw counts, parser version, dialect, and caps.

Prior workload studies commonly report structural factors such as joins, subqueries, functions, columns, and aggregations rather than a community-standard scalar complexity score~\cite{yu2018spider,li2023bird,ma2024complexsql}. SCS makes that multi-feature practice explicit, normalized, and reproducible for rewrite evaluation.

\subsection{Correctness-Gated Optimization Quality}
\label{sec:cgoq}

\CGOQ evaluates the optimization quality of a Rewritten Query $q_r$ relative to its Benchmark Input Query $q_b$. Let $T_b$ and $T_r$ denote the median runtimes of $q_b$ and $q_r$ after warm-up and five measured executions, and let $s=T_b/T_r$ be the runtime ratio. Let $C_b=\operatorname{SCS}(q_b)$ and $C_r=\operatorname{SCS}(q_r)$ be their structural scores. CGOQ is defined only when the Rewritten Query executes and satisfies the case-specific Checker Contract:
\begin{equation}
\Gamma(q_b,q_r)=\mathbf{1}\left[\begin{array}{l}
\mathrm{Executable}\land\mathrm{ResultConsistent}\land{}\\[-1pt]
\mathrm{RequiredChecksPass}
\end{array}\right].
\label{eq:correctness-gate}
\end{equation}
No-Rewrite Decisions, no-ops, non-executable outputs, and Unsafe Rewrites receive no CGOQ value; they are accounted for by the full-denominator ledger.

The runtime component uses a continuous score with a runtime-neutral band. Let $z=\ln s$, $a=\ln(1+\delta)$, and $\delta=0.05$ by default. We define
\begin{equation}
R(s)=\operatorname{sgn}(z)\tanh\left(\frac{\max(|z|-a,0)}{\tau}\right),\quad
\tau=\ln\frac{2}{1.05}.
\label{eq:runtime-score}
\end{equation}
Thus $R(s)=0$ for changes within 5\% of parity; positive values indicate speedup and negative values indicate regression. The choice of $\tau$ places the scale reference at $2\times$ speedup, where $R(2)=\tanh(1)\approx0.762$, while retaining gradual differentiation beyond that point.

Structural simplification is based on relative SCS reduction:
\begin{equation}
r_C=\max\left(0,\frac{C_b-C_r}{\max(C_b,\epsilon)}\right),
\label{eq:relative-scs-drop}
\end{equation}
where $\epsilon$ prevents division by zero. With default parameters $\theta_C=0.10$ and $c_C=0.30$, the simplification component is
\begin{equation}
K=\operatorname{clip}\left(\frac{r_C-\theta_C}{c_C-\theta_C},0,1\right).
\label{eq:simplification-score}
\end{equation}
Here $\operatorname{clip}(x,0,1)=\min(1,\max(0,x))$. SCS reductions below 10\% receive no simplification credit; reductions from 10\% to 30\% receive a linear ramp; larger reductions are capped.

To prevent structural simplification from masking a material slowdown, we define a slowdown attenuation factor. Let $\gamma=0.10$ be the slowdown cutoff at which simplification credit becomes zero, and let $[x]_+=\max(x,0)$.
\begin{equation}
B_{\gamma}(s)=1-\operatorname{clip}\left(
\frac{[\ln(1/s)-\ln(1+\delta)]_+}
{\ln(1+\gamma)-\ln(1+\delta)},0,1\right).
\label{eq:slowdown-attenuation}
\end{equation}
Thus $B_{\gamma}(s)=1$ for speedups and changes in the runtime-neutral band, decreases continuously between 5\% and 10\% slowdown, and equals zero for slowdowns of 10\% or more.

For outputs that pass the gate in Equation~\ref{eq:correctness-gate}, CGOQ is
\begin{equation}
\operatorname{CGOQ}(q_b,q_r)=100\left[R(s)+\lambda(1-|R(s)|)B_{\gamma}(s)K\right],\quad \lambda=0.4.
\label{eq:cgoq}
\end{equation}
The factor $(1-|R(s)|)$ keeps structural simplification secondary to runtime, while $B_{\gamma}(s)$ limits its ability to compensate for slowdown. Simplification receives full credit for speedups and runtime-neutral changes, partial credit for slowdowns between 5\% and 10\%, and no credit once the Rewritten Query is at least 10\% slower. Hence a large SCS reduction may offset only a small runtime regression and cannot make a materially slower rewrite positive. The score is not a proof of semantic equivalence; it is an optimization-quality value after the Checker Contract has passed.

\parhead{Parameter sensitivity.}
We evaluate $81$ configurations by varying four parameter groups: $\delta\in\{2.5,5,7.5\}\%$, $\lambda\in\{0.2,0.4,0.6\}$, the runtime scale $\tau$, and the pair $(\theta_C,c_C)$. The tested $\tau$ values correspond to reference speedups of $1.5\times$, $2\times$, and $3\times$; the tested SCS pairs are $(0.05,0.25)$, $(0.10,0.30)$, and $(0.15,0.40)$. The slowdown cutoff is fixed at $\gamma=0.10$. The Reference Rewrites stay strongly positive across the grid, with mean CGOQ between 28.54 and 57.18. All evaluated methods score below zero under every configuration. Table~\ref{tab:sensitivity} reports the score ranges; the main conclusion is stable even though lower-ranked methods exchange positions.

\begin{table}[t]
\centering
\caption{CGOQ@N ranges under the 81-configuration sensitivity grid.}
\label{tab:sensitivity}
\scriptsize
\setlength{\tabcolsep}{3.5pt}
\begin{tabular}{lrr}
\toprule
Method & Minimum & Maximum \\
\midrule
LLM-R2-plan & -0.56 & -0.52 \\
LLM-R2-queryCL & -9.71 & -6.20 \\
Kimi-K2.5 & -19.13 & -7.96 \\
LearnedRewrite & -20.57 & -14.64 \\
DeepSeek-V4 & -31.18 & -16.11 \\
R-Bot & -29.45 & -21.19 \\
GPT-5.5 & -41.46 & -20.41 \\
QUITE & -34.54 & -16.65 \\
\bottomrule
\end{tabular}
\end{table}

Rewrite-opportunity descriptions are excluded from CGOQ. They support explanation, but including them in the score would favor the Reference Rewrite's particular path over a different, equally correct and faster solution.

\subsection{Method Aggregation and Leaderboard}

Let $V_m$ be the set of Benchmark Instances for which method $m$ produces a substantive Rewritten Query that passes the Correctness Gate and obtains a CGOQ value $c_{m,i}$. We report conditional quality
\begin{equation}
\mathrm{CGOQ@Valid}(m)=\frac{1}{|V_m|}\sum_{i\in V_m}c_{m,i}
\label{eq:cgoq-valid}
\end{equation}
and full-denominator quality
\begin{equation}
\operatorname{CGOQ@N}(m)=\frac{1}{N}\sum_{i\in V_m}c_{m,i}.
\label{eq:cgoq-n}
\end{equation}
No-Rewrite Decisions, no-ops, Source Acceptance Failures, Generation Failures, Execution Failures, and Unsafe Rewrites contribute zero to Equation~\ref{eq:cgoq-n}; their counts are reported separately. A full No-Rewrite baseline therefore has CGOQ@N $=0$, which is an intentional status-quo reference rather than a loophole: if all rewrites have negative verified value, not deploying them is the safer decision. Leaderboards rank by CGOQ@N together with Generation Rate, Result Consistency Rate, UnsafeRewrite Rate, and pool-wise results, so selective behavior and unsafe activity stay visible.

\section{Evaluation Harness and Artifact}
\label{sec:harness}

\BenchName is released as an evaluation artifact at
\url{https://github.com/SQL-RewriteBench/benchmark}. The repository contains the
Executable Case Packages, baseline adapters, documentation,scripts,
and output templates needed to reproduce the benchmark workflow. The artifact
is organized to separate immutable benchmark definition from method-generated
outputs.

The \path{Case packages/} directory contains the benchmark corpus. Each
Executable Case Package stores the Benchmark Input Query, the Reference Rewrite,
the Base Query, schema metadata, provenance, result evidence,
execution-plan evidence, \path{case_info.md}, and
\path{rewrite_opportunity.md}. Baseline integrations are placed under
\path{baselines/}. User and developer documentation is placed under
\path{docs/}, including usage guides, benchmark specifications, case-package
specifications, adapter instructions, and report templates. The \path{src/} directory
contains the Evaluation Harness, command-line interface, scoring logic, and
development utilities. Evaluation outputs are written under \path{output/},
which is divided into result records, logs, and generated reports. Standard
project files such as \path{README.md}, \path{CITATION.cff},
\path{CONTRIBUTING.md}, \path{LICENSE}, and \path{pyproject.toml}
provide citation, contribution, licensing, and packaging information.

During evaluation, a method never writes into an Executable Case Package. The
Evaluation Harness reads the Benchmark Input Query and the permitted schema
context, calls a method adapter, and receives one of four normalized outcomes:
a Rewritten Query, a No-Rewrite Decision, a Source Acceptance Failure, or a
Method Error. If a Rewritten Query is returned, the Harness executes it under
the declared source dialect on the evaluation DBMS, applies the case-specific
Checker Contract, and records whether the query passes the Correctness Gate.
Only Rewritten Queries that pass the Correctness Gate proceed to runtime, SCS,
CGOQ computation.

The artifact supports two evaluation modes. In reference-pair mode, the Harness
runs the packaged Benchmark Input Query and Reference Rewrite to validate the
case and compute reference SCS and CGOQ evidence. In method-evaluation mode, the
Harness runs a baseline or user-provided adapter and computes the metric suite
for the method output. Both modes use the same Executable Case Package and
Checker Contract, which keeps reference validation and method evaluation
aligned.

The command-line interface supports the main benchmark workflow. Users can
list cases and inspect pool composition, compute SCS for Benchmark Input Queries
or Rewritten Queries, compute Pool Suitability Scores, run packaged SQL pairs,
evaluate a rewrite method through an adapter, compute CGOQ, and generate
CSV/JSON reports. The generated Rewritten Queries, execution logs, collected
plans, per-case metric records, and aggregate reports are written to
\path{output/} rather than to the case directory. This design allows users to
reproduce a leaderboard entry, inspect failures at the case level, and
regenerate tables or figures without modifying the benchmark corpus.
The artifact also freezes the complete Direct-LLM prompt, schema serialization, SQL extraction logic, raw responses, and model identifiers (DeepSeek-V4, Kimi-K2.5, and GPT-5.5), together with baseline commits and adapter configuration.

\section{Experimental Evaluation}
\label{sec:experiments}

\subsection{Setup and Methods}

Table~\ref{tab:methods} summarizes the evaluated methods and integration notes. We evaluate seven representative method families, reported as eight configurations: Direct-LLM-DeepSeek-V4, Direct-LLM-Kimi-K2.5, Direct-LLM-GPT-5.5, LearnedRewrite, LLM-R2-plan, LLM-R2-queryCL, R-Bot, and QUITE. The three Direct-LLM configurations use the same prompt and SQL-extraction logic; their model versions are DeepSeek-V4, Kimi-K2.5, and GPT-5.5. LLM-R2, R-Bot, and LearnedRewrite use GPT-5.5 in their LLM-assisted components. For LLM-R2, plan retrieves demonstrations by logical-plan tree edit distance, while queryCL retrieves demonstrations using the learned query representation released with LLM-R2 and cosine similarity. Demonstrations are drawn only from the official LLM-R2 demonstration pools, never from \BenchName Reference Rewrites.

\begin{table*}[t]
\centering
\caption{Evaluated methods and integration notes. Direct LLM methods output SQL directly; the other systems combine learned or LLM-based rule selection with a rule executor.}
\label{tab:methods}
\scriptsize
\setlength{\tabcolsep}{4pt}
\renewcommand{\arraystretch}{1.08}
\begin{tabularx}{\textwidth}{L{.16\textwidth}L{.20\textwidth}L{.22\textwidth}Y}
\toprule
Method & Selection mechanism & Rewrite executor & Integration notes \\
\midrule
DeepSeek-V4 & Prompted direct rewrite & LLM-generated SQL & DeepSeek-V4; no rule executor; failures mainly invalid or result-inconsistent SQL. \\
Kimi-K2.5 & Prompted direct rewrite & LLM-generated SQL & Kimi-K2.5; lower UnsafeRewrite Rate than DeepSeek-V4 but more non-executable outputs. \\
GPT-5.5 & Prompted direct rewrite & LLM-generated SQL & Same prompt as DeepSeek-V4 and Kimi-K2.5; high Result Consistency but many valid outputs regress. \\
LearnedRewrite & Learned rule-sequence search & Calcite-based Java service & GPT-5.5 is used in the adapter pipeline; many failures appear as service-side \texttt{Get Error}. \\
LLM-R2-plan & Logical-plan tree edit distance demo retrieval & Official Java/Calcite rule rewriter & GPT-5.5 predicts rule names; plan generation and rule execution are sensitive to PostgreSQL constructs. \\
LLM-R2-queryCL & Learned query embedding retrieval & Official Java/Calcite rule rewriter & Uses the official queryCL representation model; same downstream executor as plan mode. \\
R-Bot & Retrieval-augmented rule evidence and LLM rule ordering & Released \texttt{CalciteRewrite} module & GPT-5.5 is used for rule selection; many runs return \texttt{output\_sql=None}. \\
QUITE & Database-feedback-driven rewrite agent & Method-generated SQL & Produces high Result Consistency but often returns no-op/No-Rewrite behavior or valid regressions. \\
\bottomrule
\end{tabularx}
\end{table*}

All experiments use the frozen \NumCases release and run each executable SQL after warm-up for five measured executions, using the median runtime. Each method receives only the Benchmark Input Query and permitted schema context. A generated SQL statement is checked under the case-specific Checker Contract before SCS, speedup, or CGOQ is computed. We answer four questions: whether the packaged Reference Rewrites establish recoverable value; how current methods behave end to end; how results differ by pool and structural complexity; and why Calcite-backed rule engines produce many N/A outcomes.

\subsection{Experimental Environment}\label{sec:exp-env}

All experiments were executed on a Lenovo workstation with a 13th Gen Intel Core i7-13700 processor and 32 GB memory, running Windows 11 Pro with SQL execution inside Ubuntu 24.04.4 LTS on WSL2 and Docker 29.5.3. PostgreSQL ran in a \texttt{postgres:16} container and reported PostgreSQL 16.14. The WSL2 runtime exposed 24 logical CPUs and approximately 15 GiB memory to Docker/PostgreSQL. PostgreSQL used default-style settings, including \texttt{shared\_buffers=128MB}, \texttt{work\_mem=4MB}, and \texttt{max\_parallel\_workers\_per\_gather=2}. Each evaluated SQL query was executed after warm-up for five measured executions; the median runtime is used in all speedup and CGOQ calculations. The benchmark databases include  \texttt{dsb},  \texttt{parrot}, \texttt{sqlstorm}, and \texttt{tpcds\_sf10}, ranging from megabyte-scale parser/regression corpora to 15--16 GB analytical databases.

\subsection{Reference Rewrite Certification}

Table~\ref{tab:reference-cert} certifies the benchmark itself. All 180 Benchmark Input/Reference pairs execute and satisfy exact Result Consistency. PERF is dominated by runtime value: all 44 PERF cases obtain a positive runtime component and reach a GM Speedup Ratio of $4.12\times$. EQUIV and ROBUST mainly obtain value through SCS reduction under near-neutral runtime, while PG-AWARE exhibits both runtime and simplification value. Across all pools, the Reference Rewrites contain 91 performance-gain cases and 89 simplification-gain cases.

\begin{table}[t]
\centering
\caption{Reference Rewrite certification by frozen pool.}
\label{tab:reference-cert}
\scriptsize
\setlength{\tabcolsep}{3pt}
\begin{tabular}{lrrrrrr}
\toprule
Pool & $n$ & Perf. & Simp. & GM & Med. SCS$\downarrow$ & CGOQ \\
\midrule
EQUIV & 48 & 12 & 36 & 1.04$\times$ & 15.1\% & 15.43 \\
PERF & 44 & 44 & 0 & 4.12$\times$ & 31.1\% & 87.99 \\
ROBUST & 58 & 18 & 40 & 1.16$\times$ & 13.6\% & 20.94 \\
PG-AWARE & 30 & 17 & 13 & 1.95$\times$ & 15.7\% & 51.80 \\
\midrule
All & 180 & 91 & 89 & 1.67$\times$ & 15.9\% & 41.00 \\
\bottomrule
\end{tabular}
\end{table}

SCS is not an execution-cost model. Nevertheless, the Benchmark Input SCS has a moderate overall Spearman association with log median runtime ($\rho=0.389$, $p<10^{-7}$), while within-pool correlations are heterogeneous. This supports SCS as a structural stratification measure rather than as a replacement for measured runtime.

\subsection{Formal-Tool Coverage and Semantic-Evidence Audit}\label{sec:semantic-audit}

We also evaluated the packaged Benchmark Input/Reference pairs with SQLSolver and VeriEQL to understand how much formal evidence can be obtained for this benchmark. SQLSolver covered all 180 pairs but produced only 14 EQ results; 115 were UNKNOWN, 51 timed out. VeriEQL proved 51 pairs equivalent under its PostgreSQL setting with bound size 1, but 129 pairs returned ERROR because the tool could not parse or support the required SQL features. The two tools jointly proved 7 pairs equivalent; SQLSolver proved 7 additional pairs that VeriEQL could not decide, and VeriEQL proved 42 pairs for which SQLSolver had no EQ result.

The benchmark data are generated by TPC-DS and DSB under fixed schemas, scale factors, and workload distributions, totaling approximately 20~GB. Fact tables range from hundreds of thousands to more than 100 million rows, while dimension tables range from tens to millions. The data contain primary-key/foreign-key joins, duplicate business keys and fact rows, varied selectivities and temporal, customer, and item distributions, and actual NULL values. Result Consistency is therefore not tested on empty, single-row, duplicate-free, NULL-free, or join-inactive data. Although this evidence does not replace formal Semantic Equivalence, it exercises selections, joins, aggregations, duplicate-sensitive operations, NULL behavior, and boundary values, reducing false equivalence caused by weak validation data.

\subsection{Full-Denominator Results}

Table~\ref{tab:full-results} reports the full denominator. No method obtains positive CGOQ@N. Direct LLM methods generate on every case, but their output quality differs. GPT-5.5 achieves the strongest Result Consistency and a low UnsafeRewrite Rate, yet its CGOQ@N is negative because many result-consistent outputs regress in runtime. DeepSeek-V4 reaches high Execution Coverage but is much less safe, while Kimi-K2.5 is safer than DeepSeek-V4 but executable on fewer cases. LearnedRewrite and R-Bot lose roughly half the corpus before producing a candidate. Each LLM-R2 configuration accepts 94 of the 180 Benchmark Input Queries: 86 N/A cases terminate during mandatory source-side processing, while 69 additional accepted cases produce no final SQL. LLM-R2-plan returns 25 SQL outputs, including 18 no-ops and 7 substantive candidates; LLM-R2-queryCL returns 25 substantive candidates. QUITE achieves high Result Consistency but also regresses frequently, yielding negative CGOQ@N.

\begin{table*}[!b]
\centering
\caption{Full-denominator results over 180 Benchmark Instances. Column names match Table~\ref{tab:metrics}; percentage-valued metrics are shown with percent signs. Generation counts only non-no-op Rewritten Queries.}
\label{tab:full-results}
\scriptsize
\setlength{\tabcolsep}{3pt}
\renewcommand{\arraystretch}{1.05}
\resizebox{\textwidth}{!}{%
\begin{tabular}{lrrrrrrrr}
\toprule
Method & \makecell{Source Acceptance\\Rate $\uparrow$} & \makecell{Generation\\Rate $\uparrow$} & \makecell{No-Rewrite Decision\\Rate $\downarrow$} & \makecell{Execution Coverage\\Rate $\uparrow$} & \makecell{Result Consistency\\Rate $\uparrow$} & \makecell{UnsafeRewrite\\Rate $\downarrow$} & \makecell{CGOQ@N $\uparrow$} & \makecell{GM Speedup\\Ratio $\uparrow$} \\
\midrule
DeepSeek-V4 & 100.0\% & 100.0\% & 0.0\% & 95.0\% & 70.6\% & 24.4\% & -24.77 & 0.63$\times$ \\
Kimi-K2.5 & 100.0\% & 100.0\% & 0.0\% & 60.0\% & 53.9\% & 6.1\% & -14.72 & 0.66$\times$ \\
GPT-5.5 & 100.0\% & 100.0\% & 0.0\% & 98.9\% & 95.0\% & 3.9\% & -32.65 & 0.63$\times$ \\
LearnedRewrite & 43.9\% & 42.8\% & 1.1\% & 38.3\% & 24.4\% & 13.9\% & -18.00 & 0.33$\times$ \\
LLM-R2-plan & 52.2\% & 3.9\% & 10.0\% & 0.6\% & 0.6\% & 0.0\% & -0.55 & 0.15$\times$ \\
LLM-R2-queryCL & 52.2\% & 13.9\% & 0.0\% & 11.7\% & 11.7\% & 0.0\% & -8.09 & 0.44$\times$ \\
R-Bot & 49.4\% & 49.4\% & 0.0\% & 45.6\% & 33.3\% & 12.2\% & -25.97 & 0.33$\times$ \\
QUITE & 100.0\% & 90.6\% & 9.4\% & 89.4\% & 87.2\% & 2.2\% & -27.22 & 0.65$\times$ \\
\bottomrule
\end{tabular}}
\end{table*}

Table~\ref{tab:full-results} separates coverage, correctness, and optimization quality. High Generation Rate does not guarantee executable or safe SQL, and high Result Consistency does not guarantee positive optimization value. LLM-R2-plan has the least negative CGOQ@N largely because most cases do not reach a scored output. CGOQ@N must therefore be read with the funnel rates and terminal-status counts.

\begin{figure}[t]
\centering
\includegraphics[width=\linewidth]{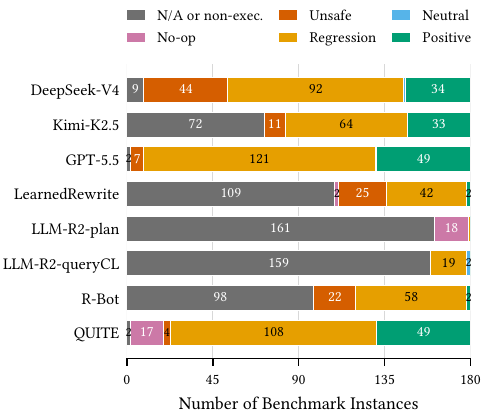}
\caption{Terminal outcome decomposition over 180 cases per method. Segment labels show absolute case counts; N/A and no-op outcomes miss value, while unsafe and regression outcomes carry active deployment risk.}
\Description{A stacked horizontal bar chart showing terminal outcomes for each evaluated method across 180 benchmark instances.}
\label{fig:outcome-decomposition}
\end{figure}

Figure~\ref{fig:outcome-decomposition} exposes why a single aggregate is insufficient. DeepSeek-V4 produces 44 Unsafe Rewrites and 92 regressions; Kimi-K2.5 has 72 non-executable outputs; GPT-5.5 has only 9 failed or unsafe outputs but 121 regressions. QUITE records 108 regressions, while LLM-R2-plan is dominated by N/A and no-op outcomes. Zero-contribution outcomes indicate missing delivered value, whereas Unsafe Rewrite and Regression indicate active risk.

\subsection{Direct-LLM Failure Analysis}
\label{sec:direct-llm-analysis}

The three Direct-LLM configurations use the same prompt, schema serialization, and SQL-extraction procedure, so their different outcomes reflect model behavior rather than a different evaluation protocol. DeepSeek-V4 executes 171 outputs, but 44 fail the Checker Contract. Among these mismatches, 37 preserve the output row count while changing values or duplicate multiplicities, and 7 change the row count. Its 9 non-executable outputs comprise 7 unresolved, ambiguous, or illegally grouped identifiers and 2 timeouts. Thus, its 95.0\% Execution Coverage does not translate into semantic safety: predicate, join, and aggregation changes can remain syntactically valid while altering the result.

Kimi-K2.5's main bottleneck is SQL completeness. Although it returns non-no-op text for all 180 cases, 72 outputs do not execute. Of these, 59 are invalid SQL, including 48 responses containing ellipses or placeholder text; the remaining 13 reach PostgreSQL but fail, with 11 caused by unresolved or ambiguous relations and columns. This explains why its Execution Coverage is 60.0\% and its Result Consistency Rate is 53.9\%, despite a comparatively low 6.1\% UnsafeRewrite Rate.

GPT-5.5 exhibits the opposite profile: 178 outputs execute and 171 pass the Checker Contract, but 121 of the result-consistent outputs are runtime regressions and only 49 obtain positive CGOQ. The regression group has a median runtime ratio of 0.438$\times$; 44 of these plans contain more physical operators than the Benchmark Input plan, and 22 introduce temporary I/O not present in the input plan. Consequently, strong executability and Result Consistency coexist with CGOQ@N $=-32.65$. The comparison shows why deployment-oriented evaluation must measure both correctness and realized benefit.

\subsection{Pool-Wise Behavior}

\begin{figure}[t]
\centering
\includegraphics[width=\linewidth]{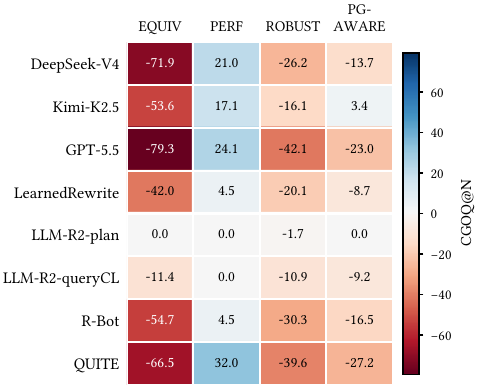}
\caption{CGOQ@N by frozen pool. Cell labels show the CGOQ@N contribution for the complete pool; non-scored cases contribute zero.}
\Description{A heatmap of CGOQ@N values by method and pool, with positive and negative values shown in contrasting colors.}
\label{fig:pool-cgoq}
\end{figure}

Figure~\ref{fig:pool-cgoq} shows that PERF is the only pool where several methods obtain positive aggregate value. QUITE recovers the most PERF value, followed by GPT-5.5, DeepSeek-V4, and Kimi-K2.5. The same systems are negative on EQUIV and ROBUST, showing that performance-oriented success does not imply robust result-preserving rewrite behavior. PG-AWARE is a same-dialect diagnostic pool for PostgreSQL-sensitive cases; it is not a cross-dialect benchmark.

\begin{table}[t]
\centering
\caption{Positive-CGOQ recovery / UnsafeRewrite Rate by frozen pool (\%).}
\label{tab:pool-recovery}
\scriptsize
\setlength{\tabcolsep}{2.5pt}
\begin{tabular}{lrrrr}
\toprule
Method & EQUIV & PERF & ROBUST & PG-AWARE \\
\midrule
DeepSeek-V4 & 4.2 / 6.2 & 38.6 / 40.9 & 12.1 / 29.3 & 26.7 / 20.0 \\
Kimi-K2.5 & 2.1 / 2.1 & 43.2 / 4.5 & 8.6 / 10.3 & 26.7 / 6.7 \\
GPT-5.5 & 2.1 / 2.1 & 63.6 / 6.8 & 20.7 / 5.2 & 26.7 / 0.0 \\
LearnedRewrite & 0.0 / 12.5 & 4.5 / 0.0 & 0.0 / 29.3 & 0.0 / 6.7 \\
LLM-R2-plan & 0.0 / 0.0 & 0.0 / 0.0 & 0.0 / 0.0 & 0.0 / 0.0 \\
LLM-R2-queryCL & 0.0 / 0.0 & 0.0 / 0.0 & 0.0 / 0.0 & 0.0 / 0.0 \\
R-Bot & 0.0 / 16.7 & 4.5 / 0.0 & 0.0 / 24.1 & 0.0 / 0.0 \\
QUITE & 4.2 / 0.0 & 68.2 / 2.3 & 19.0 / 3.4 & 20.0 / 3.3 \\
\bottomrule
\end{tabular}
\end{table}

Table~\ref{tab:pool-recovery} reports positive-CGOQ recovery together with UnsafeRewrite Rate by pool. R-Bot's positive PERF value comes from two query-41 variants. It rewrites correlated existence-style tests into grouped join structure and obtains large speedups, but it produces no broad PERF coverage. We therefore interpret this as case-level evidence rather than as pool-wide dominance. This is exactly the type of distinction a full-denominator benchmark should expose.

\subsection{Failure Modes and Complexity}

\begin{table*}[t]
\centering
\caption{Diagnosed N/A causes for Calcite-backed rule engines. Categories are derived from adapter logs and method-specific traces.}
\label{tab:na-diagnosis}
\scriptsize
\setlength{\tabcolsep}{3.5pt}
\renewcommand{\arraystretch}{1.05}
\begin{tabularx}{\textwidth}{L{.19\textwidth}rY}
\toprule
Method/configuration & N/A count & Dominant causes \\
\midrule
LLM-R2-plan & 155 & 86 source-side failures: 78 parse (52 \texttt{WITH MATERIALIZED}, 16 PostgreSQL \texttt{::} casts, 10 other syntax cases) and 8 validation (6 missing objects, 2 \texttt{substr} signatures). 69 post-acceptance N/A: 39 deterministic Calcite passes with no written SQL and 30 \texttt{no\_candidate\_sql} outcomes. \\
LLM-R2-queryCL & 155 & 86 source-side failures: 77 parse (52 \texttt{WITH MATERIALIZED}, 16 PostgreSQL \texttt{::} casts, 9 other syntax cases) and 9 validation (6 missing objects, 3 \texttt{substr} signatures). 69 post-acceptance N/A: 40 Java rule-path invocations with no final SQL and 29 deterministic Calcite passes with no written SQL. \\
R-Bot & 91 & 84 runs reached the rule-execution stage but returned \texttt{output\_sql=None}; 7 residual errors occurred after retry, mainly API-gateway failures. \\
LearnedRewrite & 101 & 99 official service errors returned \texttt{Get Error}; 2 runs returned no rewritten SQL. Schema transport succeeded, so failures primarily arise inside the parser, SQL-to-relational conversion, search, or serialization path. \\
\bottomrule
\end{tabularx}
\end{table*}

Table~\ref{tab:na-diagnosis} separates N/A outcomes by failure stage. Each LLM-R2 configuration has 86 Source Acceptance Failures and 69 post-acceptance cases without a final Rewritten Query; 140 N/A cases are shared and 15 are configuration-specific. The dominant source-side triggers are \texttt{WITH MATERIALIZED}, PostgreSQL \texttt{::} casts, other parser errors, missing catalog objects, and function-signature mismatches. Post-acceptance N/A occurs when deterministic Calcite or the selected rule path completes without a final SQL candidate. R-Bot returns N/A when Calcite yields \texttt{output\_sql=None} or API retries fail. LearnedRewrite returns N/A after internal parser, SQL-to-relational conversion, search, or serialization errors.

SCS quartiles expose another dimension of difficulty, but we interpret them as structural stratification rather than as a causal model of Source Acceptance. DeepSeek-V4's unsafe rate increases sharply from low- to high-SCS groups; GPT-5.5 keeps high Result Consistency across quartiles but still suffers from runtime regressions; LearnedRewrite and R-Bot lose coverage as queries become more nested and CTE-heavy; QUITE stays executable but continues to regress on many result-consistent outputs. Pool composition changes across quartiles, so we treat this analysis as diagnostic and rely on the pool-wise tables above for controlled comparisons. In other words, SCS helps organize failure analysis, but it is not used to claim that structural complexity alone causes Source Acceptance or optimization failure. The distinction matters because a low-SCS query can still contain a dialect feature unsupported by a front end, while a high-SCS query can be accepted when it stays within the supported grammar and relational operators.

\begin{table*}[t]
\centering
\caption{Diagnostic SCS quartile analysis. Values are Result Consistency (\%)/UnsafeRewrite Rate (\%)/CGOQ@45. Quartiles are formed by Benchmark Input SCS.}
\label{tab:scs-quartiles}
\scriptsize
\setlength{\tabcolsep}{4pt}
\renewcommand{\arraystretch}{1.08}
\begin{tabular}{lcccc}
\toprule
Method & Q1 lowest SCS & Q2 & Q3 & Q4 highest SCS \\
\midrule
DeepSeek-V4 & 84.4/11.1/-45.1 & 82.2/13.3/-38.9 & 66.7/28.9/-11.8 & 48.9/44.4/-3.2 \\
Kimi-K2.5 & 62.2/6.7/-40.0 & 62.2/4.4/-26.0 & 55.6/4.4/3.9 & 35.6/8.9/3.2 \\
GPT-5.5 & 97.8/2.2/-56.5 & 91.1/6.7/-39.5 & 95.6/2.2/-12.8 & 95.6/4.4/-21.8 \\
LearnedRewrite & 51.1/13.3/-34.2 & 28.9/15.6/-22.5 & 13.3/13.3/-11.3 & 4.4/13.3/-4.0 \\
LLM-R2-plan & 0.0/0.0/0.0 & 0.0/0.0/0.0 & 0.0/0.0/0.0 & 2.2/0.0/-2.2 \\
LLM-R2-queryCL & 28.9/0.0/-17.9 & 11.1/0.0/-8.5 & 2.2/0.0/-2.2 & 4.4/0.0/-3.8 \\
R-Bot & 68.9/8.9/-48.7 & 31.1/11.1/-24.3 & 17.8/20.0/-16.1 & 15.6/8.9/-14.8 \\
QUITE & 84.4/0.0/-45.6 & 86.7/2.2/-32.2 & 88.9/2.2/-8.5 & 88.9/4.4/-22.6 \\
\bottomrule
\end{tabular}
\end{table*}

Table~\ref{tab:scs-quartiles} shows that SCS is useful for stratifying method behavior even though it is not a front-end acceptance metric. DeepSeek-V4 becomes much less safe as structural complexity rises. GPT-5.5 sustains Result Consistency across the SCS spectrum, indicating stronger direct-generation executability, but this does not translate into positive CGOQ because many outputs are slower. LearnedRewrite and R-Bot lose result consistency from Q1 to Q4, which matches their dependence on parser, relational-conversion, and rule-execution paths. QUITE is less sensitive to SCS in Result Consistency but is still negative in CGOQ because many valid outputs regress. The quartile composition is also uneven: Q4 contains no EQUIV cases and is dominated by PERF and ROBUST. We therefore use this table as diagnostic evidence rather than as a causal claim about SCS alone.

\subsection{Calcite-Backed Rule-Engine Diagnostic}

Table~\ref{tab:calcite-slice} conditions on each Calcite-backed rule engine's accepted population. For LearnedRewrite and R-Bot, accepted cases are those for which the released implementation returns a non-N/A SQL candidate. For each LLM-R2 configuration, 86 N/A cases terminate during mandatory source-side processing; the accepted population therefore contains 94 Benchmark Instances, comprising 69 post-acceptance N/A outcomes and 25 explicit SQL outputs. LLM-R2-plan's 25 outputs include 18 no-ops and 7 substantive candidates, whereas all 25 LLM-R2-queryCL outputs are substantive. Accepted cases without valid CGOQ contribute zero.

\begin{table}[t]
\centering
\caption{Diagnostic results on the frozen Calcite-backed rule-engine slice.}
\label{tab:calcite-slice}
\scriptsize
\setlength{\tabcolsep}{3pt}
\begin{tabular}{lrrrr}
\toprule
Method & Accepted & Gen/Accept & Res. Cons/Accept & CGOQ@Accept \\
\midrule
LearnedRewrite & 79 & 97.5\% & 55.7\% & -41.02 \\
LLM-R2-plan & 94 & 7.4\% & 1.1\% & -1.06 \\
LLM-R2-queryCL & 94 & 26.6\% & 22.3\% & -15.49 \\
R-Bot & 89 & 100.0\% & 67.4\% & -52.51 \\
\bottomrule
\end{tabular}
\end{table}

This diagnostic view shows that front-end acceptance is not the only barrier. LearnedRewrite and R-Bot score strongly negative even after conditioning on their candidate-producing subset. Among the 94 accepted LLM-R2 cases, queryCL generates and validates more substantive candidates than plan, but both configurations retain negative CGOQ@Accept. Future systems therefore need broader SQL front ends, cleaner rule-executor interfaces, and result-aware selection policies rather than larger LLM prompts alone.

\subsection{Experimental Summary}

The experiments answer the four questions posed in Section~\ref{sec:experiments}. First, the packaged Reference Rewrites establish recoverable value: all 180 pairs execute, satisfy exact Result Consistency, and provide either runtime improvement or structural simplification evidence, with PERF reaching a GM Speedup Ratio of $4.12\times$ and all pools obtaining positive reference CGOQ. Second, current methods behave very differently end to end. Direct LLM methods accept all cases and generate broadly, but their failures differ: DeepSeek-V4 produces 44 Unsafe Rewrites, Kimi-K2.5 produces 72 non-executable outputs, and GPT-5.5 passes the Checker Contract on 171 cases but records 121 runtime regressions. QUITE is the most result-consistent evaluated system, but its valid outputs often slow the query. LearnedRewrite, LLM-R2, and R-Bot reveal another bottleneck: many cases fail before a deployable Rewritten Query is produced.

Third, pool-wise and SCS-stratified analyses show why a single success rate is insufficient. PERF is the only pool where several methods recover positive value, while EQUIV and ROBUST expose result-preservation and structural-coverage failures. SCS quartiles help organize these failures: some methods become less safe or less result-consistent as static structure increases, but the quartile analysis is diagnostic rather than a causal claim about SCS alone. Fourth, N/A diagnosis separates source-side rejection from post-acceptance no-output. Each LLM-R2 configuration has 86 Source Acceptance Failures and 69 additional accepted cases without a final SQL candidate; queryCL nevertheless delivers more substantive and result-consistent outputs than plan. Across the Calcite-backed methods, failures occur at parsing, validation, SQL-to-relational conversion, rule execution, SQL serialization, output extraction, and API retry stages. These findings support the main claim of the benchmark: it is not only a leaderboard, but a diagnostic instrument that points future systems toward broader SQL front ends, stronger Checker-Contract feedback, cleaner rule-executor interfaces, and more selective optimization policies.

\section{Limitations and Future Work}
\label{sec:limitations}

\parhead{Corpus growth.}
The current release contains 180 Benchmark Instances and should be read as an executable capability benchmark rather than a prevalence estimate of rewrite opportunities in production. Because \BenchName is released as an open-source project, future versions will add new Base Queries, additional Reference Rewrites, and more application-derived cases while preserving stable case identifiers and versioned release ledgers.

\parhead{Dialect and engine expansion.}
The current PG-AWARE pool is limited to PostgreSQL. Future releases will add separately executable MySQL-aware and Spark-aware tracks before making broader claims about dialect coverage. Each track will be validated on its declared source DBMS rather than through a common translation layer.

\parhead{Performance external validity.}
Runtime and speedup results are tied to the reported WSL2/Docker workstation, PostgreSQL configuration, and packaged benchmark databases. Cloud storage latency, distributed execution, caching policy, optimizer-version changes, data scale, and data skew may change speedup magnitudes. We therefore interpret current PERF results as controlled evidence that a rewrite opportunity exists, not as a universal estimate of production savings.

\section{Conclusion}
\label{sec:conclusion}

\BenchName treats statement-level SQL rewrite as an end-to-end, correctness-gated decision problem. Its \NumCases Executable Case Packages span four pools, expose documented lineage and enrichment, and pair every Benchmark Input Query with validated optimization evidence. SCS quantifies structural change, while CGOQ credits runtime improvement and gives bounded simplification credit only when runtime is neutral or the slowdown is small, after correctness is established. The Evaluation Harness preserves Source Acceptance Failures, No-Rewrite Decisions, execution failures, result mismatches, regressions, and verified gains in one denominator-explicit ledger. Across seven representative method families and eight evaluated configurations, no system achieves positive CGOQ@N; generation coverage, Result Consistency, and safety alone are insufficient for deployment.

\clearpage
\bibliographystyle{ACM-Reference-Format}
\bibliography{references}

\end{document}